\documentclass{ws-procs10x7}

\def\cM{{\cal{M}}}
\def\cP{{\cal{P}}}

\def\lam{\lambda}
\def\gam{\gamma}

\def\de{\delta}
\def\VEV#1{\left\langle#1\right\rangle}
\def\Tr{{\rm Tr}}
\def\half{\mbox{\small $\frac{1}{2}$}}
\def\ee{e^+e^-}
\def\as{\alpha_S}
\def\bas{\bar{\alpha}_S}
\def\sig{\sigma}
\def\LQCD{\Lambda_{\mbox\scriptsize QCD}}

\def\MSbar{\overline{\mbox{\scriptsize MS}}}

\begin{document}

\title{QCD Review}

\author{Giuseppe Marchesini}

\address{Universit\`a di Milano-Bicocca\\
and\\
INFN, Sezione di Milano-Bicocca, Italy}

\twocolumn[\maketitle\abstract{This review is focused on QCD
  theoretical issues and their phenomenological relevance specially
  for LHC. It is incomplete and mostly neglects the phenomenology of long
  distance physics}]

\section{Introduction}

In 1973, with the discovery of asymptotic freedom~\cite{AF}, QCD was
at the frontier of particle physics studies. Now, in 2006, QCD is
strongly installed (returning) at the center of particle physics
researches.  This, not only because of the abundance of data on jet
emission at HERA and Tevatron, but especially for the necessity of
preparing tools for discoveries at LHC. Indeed, events with large
$E_T$, as in the decay of new massive particles, are accompanied by
intense hadron emission. So the interpretation of a new elementary
process requires an accurate description of short distance hadron
physics, and this is the domain of perturbrative QCD.  Therefore
calculations have been performed in recent years in order to produce
accurate hard physics predictions for LHC.

The results of high order QCD studies go beyond their phenomenological
importance, they are exposing various new features of Feynman graphs
which may point to new general properties of quantum gauge theories.
Moreover the possibility that QCD can be viewed as a solvable string
theory is getting strength after the discovery of the AdS/CFT
correspondence.  
In the last two or three years this fact encouraged studies of
``solvable'' string theories leading to phenomenolgical results in
QCD.


QCD studies are developing in many directions and I can discuss only a
limited number of points. Before describing NLO results with their
relevance for phenomenology and for understanding new features of QCD
in general, I describe new avenues of the phenomenological attempts to
connect QCD with string theory and its impressive list of results.

\section{Phenomenology of QCD as a solvable string theory?}

String theory originated from the Veneziano model \cite{Veneziano} for
strong interactions even before the QCD era.  The possible relation of
QCD with string theory is based on various key observations.
First, QCD Feynman graphs can be embedded~\cite{thooft} on the
topological expansion of string theory (sphere, torus, etc.).
Second, a consistent string theory must have more than four
dimensions.
Recently, Polyakov \cite{Polyakov} and Maldacena \cite{Maldacena} 
suggested a holographic correspondence between a four dimensional
gauge theory and a string theory in higher dimensions.  That is, the
gauge theory observables correspond to the observables on a four
dimensional boundary of the string theory in higher dimensions.  This
correspondence was explicitly shown by Maldacena for the large $N_c$
four dimensional supersymmetric Yang-Mill theory with four
supercharges and the ten dimensional string theory with ${\rm
  AdS}_5\times S^5$ metric (AdS/CFT correspondence).
%
%
How to go from SYM theory to QCD? This question is one of the major
issues in recent string theory studies~\cite{ichep-Schomerus}, one
tries to move away from conformal symmetry and reduce the symmetries
to approach QCD.

In the last two or three years this problem has been approached also
directly form the QCD side (AdS/QCD correspondence). The question is:
what are the characteristics of a string theory that holographically
would reproduce in the four dimensional boundary the key proprieties
of QCD. Results in this direction are quite abundant. There are many
approaches with a common starting point: a string theory with a AdS
modified metric.  In five dimensions (the minimum allowed for a
possibly consistent string theory) typically one considers
$$
  ds^2\!=\!R^2\,\frac{h(z)}{z^2}\,\big(dx_\mu dx^\mu+dz^2\big)\,,
$$
with $x_\mu$ the usual four coordinates and $z$ a fifth coordinate.
The AdS$_5$ modified warping factor $h(z)$ accounts for confinement.
It breaks scale invariance introducing a scale related to the string
tension.  The QCD observables are found at the boundary $z\!=\!0$.  At
this point one tries to solve the string theory (by semiclassical
approximation) and then deduces the QCD observables typically at large
distance (the region consistent with the semiclassical approximation
used to solve the string theory). The results obtained are so numerous
that I can only list them:

\noindent
i) masses, decay width, effective couplings, chiral symmetry breaking
parameters for scalar, pseudoscalar, vector and axial
mesons~\cite{ADS-mesons}. There are few parameters to fit and the
agreement with the experimental data is reasonably good;

\noindent
ii) confinement with linear Regge trajectories and radial excitation
number~\cite{ADS-linear} . In a simple model one finds
$m_{J,n}^2=R^{-2}(J+n)$;

\noindent
iii) heavy quark potential at large and short distances; temperature
phase transition and deconfinement in agreement with lattice
calculations~\cite{ADS-AZ};

\noindent
iv) energy loss and jet quenching in heavy ion collisions~\cite{ADS-HIC};

\noindent
v) high energy scattering at fix angles and exclusive form factors
\cite{ADS-PS};

\noindent
vi) Pomeron and the BFKL anomalous dimension~\cite{ADS-BFKL}.

This long list of AdS/QCD studies speaks for itself.  String theory not
only addresses the question of unifying particle physics with
gravity, but starts to produce quite a number of phenomenological QCD
results. Whether this road is a correct one and QCD can actually be
``solved'' by string theory will be found out in the future.

\section{QCD expansion as string theory in twistor space?}

In 1986 Park and Taylor \cite{ParkeTaylor} discovered that the many
page long $gg\to gggg$ amplitude at tree level and in the maximal
helicity violation (MHV) configuration can be reduced to a small
single line formula. The formula was generalized~\cite{ParkeTaylor,BG}
to the case with any number of gluons.  The multi-gluon amplitude can
be written ($a,\lam,p=$ colour, helicity, momentum) as
$$
\cM_n=g_s^{n\!-\!2}\sum_{\rm perm} \Tr(\,t^{a_{_1}} \cdots
t^{a_{i_n}}\,)
$$
\vspace{-0.6cm}
$$
\times \,M(p_{i_1}\lam_{i_1}\cdots p_{i_n}\lam_{i_n})
$$ 
with $\Tr(\cdots)$ the colour order factors and $M(\lam_1\lam_2\cdots
\lam_n)$ the helicity and momentum ordered amplitudes.  For the MHV
configuration (all positive helicities $\lam_k\!=\!+1$ but two
$\lam_i\!=\!\lam_j\!=\!-1$) one has the single line formula
\begin{equation}
  \label{eq:mhv}
M^{\rm MHV}_{\rm tree}(1\cdots n)=
\frac{\VEV{p_ip_j}^4}{\VEV{p_1p_2}\cdots \VEV{p_np_1}}
\end{equation}
with $\VEV{p_ip_j}\!=\!\sqrt{2p_ip_j}e^{i\phi_{ij}}$ and $\phi_{ij}$
a phase proportional to the relative transverse momentum. 
This simple formula, although valid only for the MHV tree amplitudes,
indicates that Feynman diagrams have structures which are surprisingly
simple. One is then induced to find the origin of this simplicity.  An
attempt in this direction was made by Witten~\cite{Witten} who
suggested that perturbative gauge theories could be viewed as a string
theory in the twistor space. The ground for this proposal is simple. Using
the 2-spinors (massless Dirac equation solutions) one has
$$
 \VEV{pp'}\!=\!\bar u_-(p)\,u_+(p')\,,\>\> p_\mu\!=\!\half \bar
u_\pm(p)\,\gam_\mu\, u_\pm(p).
$$
This shows that, while the momentum conservation $\de^4(\sum p_i)$
depends on both $u_-$ and $u_+$ (together with the corresponding
conjugate spinors $\bar u_+\,, \bar u_-$), the amplitude $M^{\rm
  MHV}_{\rm tree}$ depends only on $u_+$ (together with its conjugate
one $\bar u_-$), see (\ref{eq:mhv}). It is then natural to perform a
``Fourier transform'' in $u_-$ and in the ``Fourier'' space, the
twistor space, one finds that the amplitude with momentum conservation
is described by a {\it line} in which the gluons are attached in an
ordered way.  The problem is then, how to describe tree amplitudes
which are not in MHV configurations and how to go beyond tree level.
The results in these two directions have been reported by
S.Moch~\cite{ichep-Moch} and Z.Berm~\cite{ichep-Bern}.

The prescription on how to go beyond MHV at tree level have been
simplified and generalized~\cite{CSW,BCFW}.  It consists of sewing
together MHV amplitudes to construct the non-MHV ones. To do this one
needs to use off-shell MHV amplitudes which are obtained by analytical
continuation in the complex momentum space.  Tree amplitudes have been
constructed for various processes: massless fermions~\cite{fermi},
Higgs boson~\cite{Higgs}, EW vector bosons~\cite{Vector}.  Results at
one loop are also obtained~\cite{BCFW}. The analytical expression for
the one-loop $gg\to gggg$ amplitude in all helicity configurations
will soon be obtained while its numerical evaluation is
available~\cite{EGZ}.

\section{High order QCD results}
Tree level amplitudes involving many QCD partons and next-to-leading
order (NLO) amplitudes are very important for LHC studies. Crucial
studies are the search for the Higgs meson and for all signals which
could indicate a way to complete the Standard model. The relevant
events involve large mass particles accompanied by intense hadron
emission. Thus it is crucial, for the interpretation of the events, to
have accurate QCD predictions. This requires calculations of a large
variety of many parton NLO matrix elements and distributions
needed for: i) direct studies of new physics signals; ii) merging
exact matrix elements with QCD resummation results for jet shape
distributions~\cite{BSZanderighi} and with Monte Carlo simulations of
QCD jet emission~\cite{MCNLO}.

There are various numerical programs to compute many-leg amplitudes
at tree level~\cite{tree-num}. The most common $t\bar t$ decay
($t\bar t\!\to\!b\bar b W^+W^-\!\to\!b\bar b q\bar q\bar q$) involves
6 final state jets.
Various processes have been computed at NLO and some presented at this
conference. The LHC ``priority'' wishlist~\cite{leshouches05} for NLO
calculations includes the
following processes\\
{\small\begin{tabular}{|l|l|} 
\hline
$V\in\{Z,W,\gamma\}$ & background to\\ 
\hline
$pp\to V\,V$\,jet & $t\bar{t}H$ \\ 
$pp\to t\bar{t}\,b\bar{b}$ & $t\bar{t}H$ \\ 
$pp\to t\bar{t}+2$\,jets& $t\bar{t}H$ \\ 
$pp\to V\,V\,b\bar{b}$ & VBF$\to H\to VV$, $t\bar{t}H$ \\
$pp\to V\,V+2$\,jets & VBF$\to H\to VV$\\ 
$pp\to V+3$\,jets & new physics signatures\\ 
$pp\to V\,V\,V$ & SUSY trilepton\\ 
\hline
\end{tabular}}
The procedure for NLO calculation of an inclusive distribution is in
principle ``simple''. Typically, one starts from the tree level
amplitude for the process under consideration with $n$-partons (Born
approximation). Then one computes the $(n\!+\!1)$-tree amplitude (real
contribution) and the one-loop correction to the Born amplitude
(virtual correction). Finally one puts together both contributions to
construct the distribution and check that collinear and infrared
divergences cancel (for regular observables).  Since in general the
inclusive sum is done numerically, the cancellation needs to be
controlled analytically first, a difficult issue which requires the
understanding of the physics of the problem (see for
instance~\cite{catani-seymour}).  Results and discussion on this
issues have been presented at this conference~\cite{NLOapproaches}.
Various techniques are used. On one hand there are seminumerical
approaches. An example is the $gg\to gggg$ one loop
amplitude~\cite{EGZ} (some of the helicity configurations have been
computed analytically by twistor techniques). On the other hand there
are direct analytical approaches. Powerful methods~\cite{ichep-Bern}
are based on Cutkosky rules (unitarity), on the use of
Passarino-Veltman~\cite{PV} reduction of any one-loop amplitudes in
terms of a basis of scalar integrals and on recurrence relations.

The results of these studies are usually obtained after profound
understanding of the general structure of Feynman diagrams. Often one
find general properties in gauge theories which points toward simple
structures.

Contributions to this conferences on phenomenological studies at Lep,
Hera, Tevatron and LHC of high order results results have been
presented. They are:

\noindent
i) running coupling measurements~\cite{ichep-as} at Hera and Lep.
Lattice~\cite{ichep-Schierholz} calculations can be used to reduce the
theoretical errors;

\noindent
ii) jet emission studies~\cite{ichep-jets} at Lep, Hera, Tevatron and
LHC. In particular at the Tevatron $k_t$ jet-finding algorithms start
to be used;

\noindent
ii) heavy flavour production~\cite{ichep-heavy} at Hera and Tevatron;

\noindent
iii) W/Z and W/Z+jets at Tevatron~\cite{ichep-WZ};

\noindent
iv) Higgs and W/Z production to NNLO~\cite{ichep-Higgs} at LHC.

The general comment for these analyses is that NLO corrections improve
the accuracy and the description of the data.

\section{High order parton splitting}
Parton density functions enter DIS and, due to QCD factorization,
hadron collider distributions.  Fragmentation functions, which describe
inclusive final state emission, enter all collider studies. Their
$Q^2$-evolution is governed by the corresponding space- and time-like
parton splitting functions (anomalous dimensions) which have been
computed~\cite{CFP} in 1980 at two loops both for the singlet and
non-singlet cases. Recently the anomalous dimensions have been
computed at three loops: the singlet and non-singlet ones for the
space-like case~\cite{MVV}; the non-singlet ones for the time-like
case~\cite{MMV}. These very important results obtained in $\MSbar$
scheme have been already used for various phenomenological
studies~\cite{vari-VVMM}: Sudakov resummations, lepton pair
and Higgs boson production, quark form factor, threshold resummation,
DIS by photon exchange, longitudinal structure function, non-singlet
analysis of deep inelastic world data.

High order anomalous dimensions are also important to understand
general features of QCD and gauge theories in general. I discuss here
two examples: relation between DGLAP and BFLK evolution in SYM
theory and relation between space- and time-like anomalous dimensions.

{\bf DGLAP and BFLK evolutions}.  It has been
shown~\cite{ichep-koticov} that in the ${\mathcal N}=4$ Supersymmetric
Yang-Mills theory there is a deep relation between the BFKL and DGLAP
evolution equations.  In this theory, the eigenvalues of the
space-like anomalous dimension matrix are expressed in terms of a
universal function constrained (obtained) from the BFKL equation.
This was checked at two loops by direct calculations   
and at three loop with the anomalous dimensions obtained for SYM from
the QCD ones~\cite{MVV}.  It is important to explore to what extent
the relations between BLKF and DGLAP can be extended to QCD.

{\bf Relating S- and T-evolution}.  The search for a relation between
S- and T-anomalous dimension (space-like $\gam_-(N)$ and time-like
$\gam_+(N)$) has a long story: Drell-Levi-Yan relation~\cite{DLY},
Gribov-Lipatov relation~\cite{GL}, the analytical
continuation~\cite{CFP,StratmannV}.
Consider DIS with $q$ the large space-like momentum transferred from
the incident lepton to the target nucleon $P$ and $\ee$ annihilation
with $q$ the time-like total incoming momentum and $P$ the final
observed hadron.  The Bjorken and Feynman variables in DIS and $\ee$
are
$$
  x_B \>=\> \frac{-q^2}{2(Pq)}, \qquad  x_F \>=\> \frac{2(Pq)}{q^2}.
$$
These variables are mutually reciprocal: after the crossing operation
$P\to -P$ one $x$ becomes the inverse of the other (although in both
channels $0\le x \le 1$ thus requiring the analytical continuation).
This fact was the basis for the search of reciprocity relations
between $\gam_-(N)$ and $\gam_+(N)$ (here $N$ is the Mellin moment
conjugate to $x$).

Recently it has been noticed that new information on the relation
between $\gam_-(N)$ and $\gam_+(N)$ could be obtained by taking into
account that such a reciprocity property $x\!\to\!1/x$ ($x\!=\!x_B$ or
$x_F$) can be extended to the Feynman diagram for the two processes
and, in particular, to the contributions from mass-singularities
described by multi-parton splitting.
Consider the three-parton vertex kinematics of the decay $k_0 \to k +
k'$ in the DIS situation: $k_0^2<0$, $k^2<0$, ${k'}^2>0$.  To change
to the annihilation kinematics, $-k \to -k_0 + k'$, one has to change
signs of $k_{+0}$ and $k_+$ and of the corresponding virtualities,
$k_0^2>0$, $k^2>0$.
The virtuality $k^2$ enters the denominators of the Feynman diagrams.
In order for the transverse momentum integration produce a logarithmic
enhancement, the conditions must be satisfied
\begin{equation}
\label{eq:rrkin}
|k_0^2| < \frac{k_{+0}}{k_+}|k^2| = z^\sig\cdot \kappa^2,\qquad \sigma=\pm  
\end{equation}
with $\sig\!=\!-1$ and $\sig\!=\!1$ for S- and T-case respectively.
This kinematical fact has a strong impact on the relation between the
S-and T-probability $D_\sig(N,\kappa^2)$ to find a parton with
virtuality up to $\kappa^2$. From (\ref{eq:rrkin}) one directly
deduces the following {\it reciprocity respecting}
equation~\cite{DokshitzerMS} (RRE)
\begin{eqnarray}
\label{eq:rre}
&\kappa^2&\partial_{\kappa^2}D_\sig(N,\kappa^2)= 
\gam_\sig(N)\, D_\sig(N,\kappa^2)\\ 
&&=\int_0^1\frac{dz}{z}\,z^N\,P(z,\as)\,
D_\sig(N,z^\sig\kappa^2)\,.\nonumber
\end{eqnarray}
The difference between the two channels is simply in the fact that the
virtuality of the integrated parton distribution is
$\kappa^2\,z^\sig$, see (\ref{eq:rrkin}). The splitting function
$P(z,\as)$ does not depend on the S- or T-channel (its Mellin moments
are not the anomalous dimensions).  The running coupling in the
splitting function depends~\cite{DokshitzerMS} on the virtuality in a
reciprocity respecting form.
This equation (in general a matrix equation) is non-local: for
$\sig\!=\!-1$ ($\sig\!=\!1$) the right hand side involves the parton
distribution with all virtualities larger (smaller) than $\kappa^2$.  So
RRE is not suitable for explicit calculations of the anomalous
dimensions, but for relating them.

Is eq.~(\ref{eq:rre}) correct?  It is the result of the vertex
kinematical ordering (\ref{eq:rrkin}) for mass singularities.
However, when dimensional regularization is used, the implication of
the vertex kinematical ordering gets mixed with the fact that the S-
and T-channel phase space differ by a factor $z^{-2\epsilon}$.
However corrections coming from this $z$-factor do not lead to really
new structures but are essentially related to the anomalous dimensions
at lower orders.  This suggests (see discussion in \cite{StratmannV})
that these corrections could be an artifact of dimensional
regularization so that, at the end of the calculation, reciprocity is
actually restored leaving RRE unmodified.

The reciprocity relation (\ref{eq:rre}) can be tested for higher order
S- and T-anomalous dimensions~\cite{MVV,MMV}.  To do that one has to
account also for the arguments of the running coupling in S- and
T-cases which give contributions proportional to the beta-function.
However beta-function contributions arise also from the factorization
scheme used for S- and T-case and how to account for their reciprocity
has not been studied yet.  Then RRE have been  tested only for the fixed
$\as$ contribution. In this case RRE can be written as
\begin{eqnarray}
\label{eq:rr}
&\gam_\sig(N)& = \cP(N+\sig\gam_\sig),\\
&&\cP(N)=\int_0^1\frac{dz}{z}z^N\,P(z)\nonumber\,,
\end{eqnarray}
with $\cP(N)$ a universal function depending on $\as$.
Eq.~(\ref{eq:rr}) has been tested to high order in three cases: the
non-singlet case, the large $x$ and the small $x$ behaviour.

\noindent
1) {\it Non-single case}. From (\ref{eq:rr}) one has
\begin{eqnarray}
\gam^{1}_\sig(N)=&&\cP^{1}(N)\nonumber\\
\gam^{2}_\sig(N)=&&\cP^{2}(N)+\sig\gam^1(N)\,\dot\gam^{1}(N)\nonumber\\
\gam^{3}_\sig(N)=&&\cP^{3}(N)+\half(\gam^{1}(N))^2\,\ddot\gam^1(N)\nonumber\\
&&+\sig(\gam_\sig^2(N)\,\dot\gam^{1}(N)+\gam^{1}(N)\dot\cP^{2}(N)\nonumber
\end{eqnarray}
with $\gam_\sig^n$ and $\cP^n$ the $n$-th expansion coefficients in
$\as$ of non-singlet $\gam_\sig(N)$ and $\cP_\sig(N)$. Dots are
derivatives with respect to $N$.  The first is the Gribov-Lipatov
relation (independence of $\sig$ valid only to one loop). The second
is the two loop relation which has been pointed out in~\cite{CFP} and
has been one of the important elements used~\cite{DokshitzerMS} to
derive RRE. The last one has been verified~\cite{MMV} at three loop
order.

\noindent 
2) {\it Large $x$ behaviour}. The dominant channels are the
diagonal ones $gg$ and $qq$. Denoting by
$\tilde\gam_\sig^{(a)}(x)$ the $x$-space anomalous dimension with
$a=gg$ or $q\bar q$, one has
\begin{eqnarray}
\tilde\gam^{(a)}_\sig(x)=&&\frac{A^{(a)}\,x}{(1\!-\!x)_+}+
B^{(a)}\delta(1\!-\!x)\label{eq:as-phys}\\
&&+C_\sig^{(a)}\ln(1\!-\!x)+D_\sig^{(a)}+\cdots\nonumber
\end{eqnarray}
The various coefficients are functions of $\as$.  The two most
singular terms do not depends on $\sig$. In particular the first term
corresponds to the classical soft radiation~\cite{LBK} which is
universal and depends only on the charge $A^{(a)}$ of the source. It
can by expressed in terms of a physical coupling~\cite{asphys} as
$A^{(a)}=(C_a/\pi)\as^{\rm phys}$ with $C_a=C_A, C_F$ for $=gg,q\bar q$. 
Using (\ref{eq:rr}) one deduces 
$$
C_\sig^{(a)}=-\sig(A^{(a)})^2\,,\quad
D_\sig^{(a)}=-\sig A^{(a)}\,B^{(a)}\,.
$$
These two relations are verified at three loop level, apart for a
beta-function contribution entering $D_\sig^{(a)}$ which is not
considered in (\ref{eq:rr}). 

Very recently the expansion of the leading coefficient $A$ has been
evaluated~\cite{ES} in SYM with ${\cal{N}}\!=\!4$ to all order in
$\as$ (in dimensional regularization scheme suited for supersymmetric
theories) in the S-case. All coefficients satisfy the
``trascendentality principle'' (i.e. are given in terms of derivative
of the Euler psi-function). The first three terms agree with the
calculation reported in~\cite{ichep-koticov}. According to
(\ref{eq:as-phys}) the same result should be obtained for the T-case.

\noindent 
3) {\it Small $x$ behaviour}. The leading order contributions for
$N\!\to\!0$ (corresponding to small $x$) are given by
($\bas=C_A\as/\pi$)
\begin{eqnarray}
\gam_-(N)\!=\!\frac{\bas}{N}\,,\>\>\>
\gam_+(N)\!=\!\frac14(\sqrt{N^2\!+\!8\bas}-\!N)\nonumber
\end{eqnarray}
These two expressions satisfy (\ref{eq:rr}). They are the result of
cancellations in the phase space due to coherence of soft radiation:
angular ordering for the T-case and transverse momentum ordering for
the S-case. Thus cancellations in the mass singularity phase space are
reciprocity related so that RRE incorporates them into the universal
function $P(z)$. RRE can be tested to higher order. One of the most
interesting outputs is that the ``accidental'' absence in the leading
BFKL anomalous dimension of the $\as^2/N^2$ and $\as^3/N^3$ terms
implies, via reciprocity (\ref{eq:rr}), the fact that exact angular
ordering is valid beyond leading order up to NNLO.

\section{Additional problems in hadron-hadron collisions}

Hard processes in hadron collider are initiated by elementary hard
cross sections with 2 incoming and n outgoing partons
\begin{eqnarray}
&p_1p_2\to 0\quad& \mbox{DY, WW, ZZ }\cdots\nonumber\\
&p_1p_2\to 1\quad& p_t{\rm -Higgs}, \mbox{Higgs+jet}\cdots \nonumber\\
&p_1p_2\to 2\quad& \mbox{dijet-distributions, jet-shape}\cdots\nonumber\\
&p_1p_2\to n\quad& n>2\,,\>\> \mbox{multi-jet distributions} \nonumber
\end{eqnarray}
For $n\!\ge\!2$ one has a new QCD challenge.  Consider inclusive
distributions with two different hard scales
($Q\!\gg\!Q_0\!\gg\!\LQCD$) for which logarithmic resummations are
needed. An example is the out-of-event-plane energy distribution.
Because of the complex structure of colour matrices for
$n\!\ge\!2$, soft gluon at large angles contribute to single
logarithmic accuracy. Consider the hard vertex with four partons
\begin{equation}
  \label{eq:2to2}
p_1\,p_2\to p_3\,p_4
\end{equation}
Resummation of collinear and infrared logarithmic contributions coming
from radiation emitted off the four primary QCD partons gives rise to
four Sudakov form factors. Each factor has the charge of the emitting
parton and the hard scale identified (as in $e^+e^-$) by the single
logarithmic analysis. To single log accuracy one has additional
contributions coming from large angle soft QCD emission which are
resummed into a fifth form factor\cite {fifth}. To understand why
these large angle soft terms enter only for $n\ge2$ consider the hard
process (\ref{eq:2to2}) with matrix colour charges $T_i$. Emission of
a soft gluon $q$ off these four partons is given by the square of the
eikonal current. Using charge conservation
($T_1\!+\!T_2\!=\!T_3\!+\!T_4$) one has
\begin{eqnarray}
&j^2(q)=(\sum_iT_i\frac{p_i}{p_iq})^2\label{eq:j2}\\
&=\!\sum_i T^2_i\,W_i(q)+ T_{t}^2\,A_t(q)+T_{u}^2\,A_u(q)\nonumber
\end{eqnarray}
with $T_i^2\!=\!C_i$ the square colour charge and $W_i(q)$ the
collinear and infrared divergent emission factor leading to the
standard Sudakov form factor of the primary parton $i$.  The emission
factors $A_s$ and $A_t$ are infrared but not collinear singular.
While $T_i^2$ are proportional to unity, the exchanged charges
$T_t^2\!=\!(T_1\!-\!T_3)^2$ and $T_u^2\!=\!(T_1\!-\!T_4)^2$ are colour
matrices ($6\times6$ in $SU(N)$ for $gg\!\to\!gg$).  These last two
terms, together with the corresponding Coulomb phases, give rise to
the matrix fifth form factor (which needs to be diagonalized). From
this discussion it is clear that these additional soft, but non
collinear, contributions are absent for less than four colour
particles since there isn't any exchanged channel.

For $gg\to gg$ there is a puzzle.  The relevant eigenvalues of the
soft distribution (\ref{eq:j2}) are symmetric under the exchange
between external and internal space variables:
$$
\frac{\ln\frac{s^2}{tu}\!-\!2\pi i}{\ln\frac{t}{u}} \quad 
\Longleftrightarrow \quad N_c\,.
$$
Another surprising result concerns the contribution beyond single-log
(next-to-next-to-leading) to the fifth form factor: it has been
found~\cite{ADSterman} that it is proportional to the corresponding
one-loop contribution and again one reconstructs the physical coupling
$A^{(a)}$ in (\ref{eq:as-phys}).  This result makes possible a variety
of resummations at next-to-next-to leading order.

\section{Jets, small-$x$ and all that}

\noindent {\it Jets shape distributions}.  Examples in hadron hadron
collisions are the out-of-event-plane energy distribution and the
energy-energy azimuthal correlations. Both reliable predictions and
experimental data~\cite{ichep-Levonian,ichep-Kupco} are difficult to
obtain. The large number of hadrons emitted at LHC adds a further
algorithmic difficulty.  A typical $k_t$ jet-finder algorithms scales
as $n^3$ with $n$ the number of particles. A good news is that a
jet-finder algorithms was found~\cite{CacciariS}, using techniques
developed in computational geometry, that scales as $n \ln n$. The
techniques required to obtain reliable jet distribution predictions
are well understood: single logarithmic resummation, matching with
known fix-order NLO results and non-perturbative power corrections.
Reliable predictions have been obtained when no more than four jet are
involved, counting also incoming jets that is one in DIS and two in
hardon-hadron collisions. In this last case most events have four
jets and analytical calculations become very laborious (see previous
section) and the use of automate resummation~\cite{BSZanderighi}
becomes essentially unavoidable.  An additional difficulty in the
study of hadron emission associated to new physics events is that one
may want to describe QCD radiation in given geometrical regions. In
such a case one needs~\cite{DasguptaS} to resum non-global logs and
this requires solving non-linear equations.

\noindent {\it Small-$x$ physics} enters a large number of processes
and gives rise to many interesting problems. The connection between
BFKL and DGLAP equations has been already mentioned.  Important
developments are on {\it linear} and {\it non-linear} small-$x$
regimes.

In the linear regime there have been extensive studies of the
higher-order corrections which could stabilize the NLO poor
perturbative convergence and construct a framework useful for
phenomenological applications. The general method
consists~\cite{smallxRG} of matching small-$x$ resummation with
collinear singularity resummations (running $\as$, anomalous
dimensions, factorization scheme...).

The non-linear regime, a major issue in small-$x$ physics, is
characterized by saturation~\cite{ichep-Marquet} which should be seen
in heavy ion collisions~\cite{HIC}. A major point is how to go beyond
the Balitsky-Kovchegov equation viewed as a mean field approximation
of a more general equation taking into account unitarity in
perturbative QCD. A number of new formulations have been proposed, but
it seems that a complete and consistent formulation is still lacking.
An important attempt to account for general non-linear QCD corrections
at small-$x$ is based on the observation~\cite{MPesch} that the BK
equation is in the same universality class as the
Fisher-Kolmogorov-Petrovsky-Piscounov equation introduced in
statistical physics to discuss reaction-diffusion phenomena. Therefore
methods of stochastic physics are used~\cite{stochastic} to study
small-$x$ QCD phenomena (and viceversa).

Perturbative QCD studies~\cite{ichep-bartels} of multiple interactions
in DIS and hadron-hadron collisions are important for various reasons:
they are needed to set up the theoretical framework for the small-$x$
non-linear equations as the BK equation and to perform
phenomenological studies including multi-jet final states (background
to new physics), heavy flavour jets (near forward direction) and
underlying event~\cite{ichep-field}.

Diffraction is phenomenologically important since even hard events have
diffractive components. Moreover, diffraction could be used to study
Higgs production~\cite{ichep-forshaw}. Diffraction is a
non-perturbative fields and phenomenological models are needed. Their
basis is unitarity and interactions  between Pomerons~\cite{ichep-kaidalov}.

\section{Final remarks}
This (theoretical) review of QCD is incomplete and mostly neglects the
phenomenology of long distance physics. The list of issues here
discussed
and of the ones not discussed (NLO matching with Monte Carlo
simulations, PDF and structure functions, photon emission, two photon
scattering, prompt photons, power corrections, Bjorken and
Gross-Llewellyn Smith sum rules, underlying events...)  speaks for
itself of the importance of QCD for high energy physics in general.
QCD for LHC physics poses new problems and gives new information on
emission such as rotation in colour space and consequent non-planar
corrections.

Very laborious calculations, needed for LHC accurate predictions,
reveal simple structures and properties. This fact points to the
possibility of a new more efficient formulation of QCD. Is it possible
that this could be provided by ideas developed in string theory? There
are indications in this directions although not yet convincing: the
abundant phenomenological results obtained within the framework of
AdS/CFT as a way to ``solve'' QCD; Feynman diagrams in twistor
formulations.  However, for the moment, all these developments are
only formal and, to make a decisive step in the understanding of
QCD, one needs to find their physics basis.

\end{document}